\documentclass[lineno]{jfm}

\usepackage{graphicx}
\usepackage{epstopdf,epsfig}
\usepackage{newtxtext}
\usepackage{newtxmath}
\usepackage{makecell}
\usepackage{natbib}
\usepackage{hyperref}

\newcommand\Oh{\mbox{\textit{Oh}}}

\newcommand{\RomanNumeralCaps}[1]
\linenumbers
\usepackage{soul,color,xcolor}
\sethlcolor{yellow}
\soulregister{\cite}7
\soulregister{\citep}7
\soulregister{\citet}7
\soulregister{\ref}7
\soulregister{\pageref}7

\title{Coalescence of immiscible sessile droplets on a partial wetting surface}

\author{Huadan Xu,
Xinjin Ge,
  Tianyou Wang
 \and Zhizhao Che
\corresp{\email{chezhizhao@tju.edu.cn}}}
\affiliation{State Key Laboratory of Engines, Tianjin University, Tianjin 300350, China
}

\begin{document}
\maketitle

\begin{abstract}
Droplet coalescence is a common phenomenon and plays an important role in multi-disciplinary applications. Previous studies mainly consider the coalescence of miscible liquid, even though the coalescence of immiscible droplets on a solid surface is a common process. In this study, we explore the coalescence of two immiscible droplets on a partial wetting surface experimentally and theoretically. We find that the coalescence process can be divided into three stages based on the timescales and force interactions involved, namely (I) the growth of the liquid bridge, (II) the oscillation of the coalescing sessile droplet, and (III) the formation of a partially-engulfed compound sessile droplet and the subsequent retraction. In stage I, the immiscible interface is found not to affect the scaling of the temporal evolution of the liquid bridge, which follows the same 2/3 power law as that of miscible droplets. In Stage II, by developing a new capillary timescale considering both surface and interfacial tensions, we show that the interfacial tension between the two immiscible liquids functions as a nonnegligible resistance to the oscillation which decreases the oscillation periods. In Stage III, a modified Ohnesorge number is developed to characterize the visco-capillary and inertia-capillary timescales involved during the displacement of water by oil; a new model based on energy balance is proposed to analyze the maximum retraction velocity, highlighting that the viscous resistance is concentrated in a region close to the contact line.
\end{abstract}

\section{Introduction}\label{sec:1}
Droplet coalescence has long been an important research topic and attracted growing interest due to its close relevance to broad applications such as material synthesis \citep{Sohrabi2020DropletsMicrofluidics, Song2006MicrofluidicChannels}, aircraft anti-icing \citep{Johannes2017technicalLiquid, Cha2016Antiice}, and efficient cooling \citep{Gong2017CondensateMicrodrop, Wang2022DropletWetting}. Most of the previous studies of droplet coalescence considered the coalescence of two spherical droplets \citep{Duchemin2003InviscidCoalescence, Paulsen2013DropCoalescence, Thoroddsen2007MiscibleDrople}. The coalescence dynamics of two sessile droplets (i.e., liquid droplets sitting on solid substrates), in comparison, is inherently different due to the presence of a solid substrate \citep{Eddi2013DropletGeometry, Hernandez2012Coalescence, Pawar2019CoalescenceSolid, Sui2013InertialCoalescence}. Considering its great importance in applications in digital microfluidics \citep{Kihwan2012DigitalMicrofluidics}, precise control of ink-jet printing \citep{Lee2013Printing}, and dropwise condensation \citep{WANG2022dropletCondensation}, the coalescence of sessile droplet continues to be an active area of research. The coalescence of sessile droplets could be strongly affected due to the interaction with solids surfaces \citep{Neogi1982SpreadingSolid}. Due to the extra resistance from the solid surfaces, the coalescence of sessile droplets consists of other different dynamics, such as oscillation \citep{Jiang2019SessileMicrodroplet, Somwanshi2018PendentDroplet} and contact line relaxation \citep{Beysens2006ContactLine, Gokhale2004CondensingDroplet}.

When two spherical droplets are made to touch and coalesce, a small liquid bridge forms connecting the two droplets driven by capillary forces \citep{Aarts2005dropletCoalescence, Thoroddsen2007MiscibleDrople, Xu2022BridgeEvolution}. Then depending on the competition between inertia and viscous forces, the radius of the liquid bridge $r$ grows as $r\sim t$ (if viscous force is the dominant resistance) or as $r\sim {{t}^{{1}/{2}}}$ (if inertia is the dominant resistance), with a crossover length depending on the size of droplets and fluid properties \citep{Paulsen2013DropCoalescence}. However, the bridge dynamics of two sessile droplets could show radically different behaviors compared to that of two spherical droplets due to their different geometries caused by the presence of substrate. Recently, the influence of geometry (the initial shape of the droplet before the coalescence) on the coalescence dynamics was investigated. The short-time dynamics of droplets coalescence on solid surfaces can be characterised by the liquid bridge evolution \citep{Eddi2013DropletGeometry, Pawar2019CoalescenceSolid, Sui2013InertialCoalescence} and the growth of the bridge height ($h_m$) follows a power law ${{h}_{{m}}}\sim {{t}^{\alpha }}$. For viscous droplets on a partial wetting surface, the time evolution of the liquid bridge was found to grow linearly with time (i.e., $\alpha = 1$), where the droplet was introduced either by liquid deposition or vapour condensation \citep{Hernandez2012Coalescence, Narhe2008ContactLine}. In another relevant study where silicone oil droplets were put on a perfectly wetting substrate, the droplet's height and radius before the contact were demonstrated to have a large influence on the growth rate of the liquid bridge \citep{Ristenpart2006SpreadingDroplet}. Lee \emph{et al.} (\citeyear{Lee2013Printing}) conducted an experiment on wettable surfaces with contact angles ranging from 10$^\circ$ to 56$^\circ$ using an electrohydrodynamic (EHD) inkjet system and obtained a power-law exponent of $0.51\le \alpha \le 0.86$. In another investigation by Eddi \emph{et al.} (\citeyear{Eddi2013DropletGeometry}) using two equal-sized water droplets, they discovered that the liquid bridge dynamics is related to the initial shape of the droplet, following an exponent of 2/3 for a contact angle smaller than 90$^\circ$, and an exponent of 1/2 when the contact angle approaches 90$^\circ$ \citep{Eddi2013DropletGeometry}.

After the completion of the liquid bridge growth, there follows a slower process involving the oscillation and the relaxation of the coalescing droplet. As the presence of a substrate could introduce extra resistance \citep{Neogi1982SpreadingSolid}, different oscillation dynamics are expected compared with the coalescence of spherical droplets. Somwanshi \emph{et al.} (\citeyear{Somwanshi2018PendentDroplet}) did experiments on the coalescence of two droplets attached to a hydrophobic surface either in pendent or sessile mode, and they found that the wall shear stress at the solid surface is much larger in the sessile mode than in the pendent mode. Jiang \emph{et al.} (\citeyear{Jiang2019SessileMicrodroplet}) studied the coalescence of two droplets with various wettability and found the capillary wave oscillation gets weaker with the decrease of the surface wettability. In a longer timescale, droplet dynamics is resisted by the viscous resistance from the bulk fluid or from the contact line. This stage could take up to several seconds depending on properties of the liquid and the substrate \citep{Jiang2019SessileMicrodroplet}. Some other phenomena would arise in this process, such as contact angle variations and contact line movements. For example, the experiments on droplet coalescence on partial wetting surfaces conducted by Narhe \emph{et al.} (\citeyear{Narhe2004ContactLineSpreading}) and Andrieu \emph{et al.} (\citeyear{Andrieu2002SessileCoalescence}) showed that in the late stage, the droplet slowly relaxes to a circular shape which lasts six or seven orders of magnitude longer than that described by the bulk hydrodynamics. Such slow relaxation was demonstrated to be induced by the liquid-vapour phase change near the contact line. Zhang \emph{et al.} (\citeyear{Zhang2015GravityCurrent}) observed complete stratification of the liquids of two droplets with different densities and viscosities followed by the diffusive mixing in the composite droplet over a timescale of several minutes. For the coalescence of an impacting droplet with a sessile droplet, the induced surface jet could be either promoted or suppressed by the Marangoni flow, depending on the direction of the surface tension difference \citep{Sykes2020SurfaceJet}.

Most of the previous efforts on the coalescence of sessile droplets are restricted to miscible droplets \citep{Borcia2013PartialCoalescence, huang2021transitions, ahmadlouydarab2014motion}. For the limited studies reporting directly or indirectly immiscible sessile droplets interaction, many focused on the equilibrium state after the interactions between the immiscible droplets, in which they either investigated the role of interfacial energy in forming a stable compound structure \citep{Mahadevan2002FourPhaseMerging}, or they considered the effect of immiscible interface on a temporary equilibrium of a smaller oil droplet on a bigger water droplet \citep{Iqbal2017CompoundDroplet}. Rostami and Auernhammer (\citeyear{Rostami2022CapillaryMerging}) analyzed the motion of the four-phase point (where the two liquids, the gas, and the solid meet). From the perspective of the liquid spreading in a V-shape groove, they showed that the dynamics of this point is independent of contact line velocity but rather similar to the capillary flow in a tube. Another related work is by Xu \emph{et al.} (\citeyear{Xu2022BridgeEvolution}), who experimentally and theoretically investigated the liquid bridge dynamics of a pendent immiscible droplet with a sessile droplet, and found that the immiscibility of the two liquids results in slower growth of the liquid bridge. These results illustrate the potential influence of miscibility and liquid wettability on immiscible droplet systems. However, when it comes to the coalescence of immiscible droplets on a substrate, the potential influences are expected to be much more complicated as they not only hold an extra immiscible interface \citep{de2004capillarity}, but also keep respective surface tensions and different interfacial tensions with the solid surface throughout the coalescence process. Thus, these different characteristics might have significant influences on the short-time dynamics of coalescing immiscible droplets, as well as their subsequent long-time behaviours of the coalesced droplets. However, despite its wide applications in multiphase processing \citep{Xie2022Demulsification}, material synthesis \citep{Winkelmann2013miniemulsions}, and biological interaction of cells \citep{Kusumaatmaja2021IntracellularWetting}, a clear understanding of the coalescence of immiscible sessile droplets is still missing.

Based on the understanding of the previous research on droplet coalescence, we experimentally investigate the coalescence dynamics of two immiscible droplets sitting on wettable surfaces. We identify three stages of coalescence, and analyze them in detail. We first study the fast growth of the liquid bridge after the coalescence and consider low-viscosity and high-viscosity oil droplets separately. We then analyze the oscillation dynamics of the coalescing droplet and identify the role of the water-oil interfacial tension. Finally, the retraction of the coalesced droplet in a longer timescale is investigated, and the retraction dynamics for various immiscible droplet pairs with different surface tensions, interfacial tensions, and viscosities are quantified.

\section{Experimental method}\label{sec:2}
\begin{figure}
  \centering
  \includegraphics[scale=0.9]{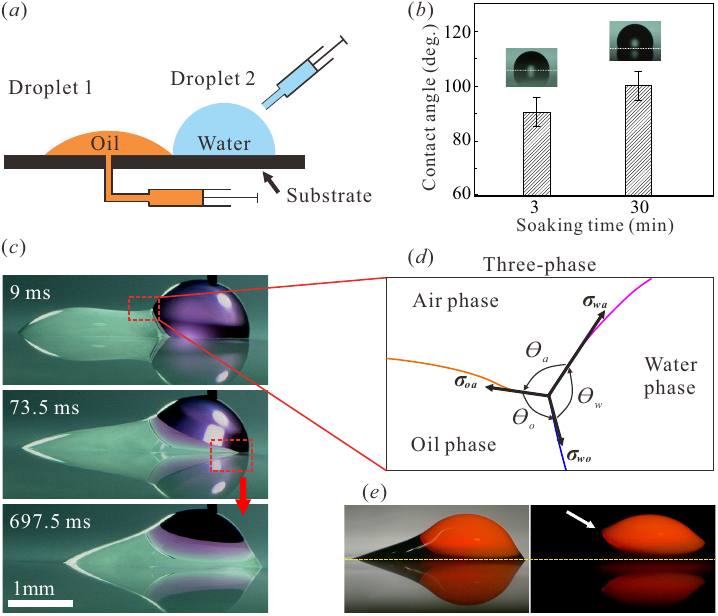}
  \caption{(a) Schematic diagram of the experimental setup for the coalescence of two sessile immiscible droplets. (b) Contact angle of the substrates under two different soaking times in the hydrophobic solution. (c) Typical snapshots during the coalescence of the two immiscible droplets. The water droplet was dyed red by Rhodamine B with a concentration lower than 0.1 wt\%. (The needle embedded in the water droplet was used (here only) to deposit the water droplet on the substrate, image used here only to give a qualitative view of the four-phase capillary driven flow in the contact region) (d) Sketch of three liquid interfaces in the contact region, where the three-phase contact line generates three contact angles (${\theta }_{a}$, ${\theta }_{w}$, and ${\theta }_{o}$). (e) Illustration of the dyed water droplet being displaced from the substrate by using different lighting methods: image to show the compound droplet with simultaneous front lighting and back lighting (left), and image to show the dyed water droplet alone with only front lighting (right). The dashed yellow line denotes the position of the substrate. The dyed water droplet is a mixture of 50 wt\% glycerol solution with a low concentration of Congo red.
  }\label{fig:01}
\end{figure}

   In the experiment, a water droplet with a known volume was first deposited on the substrate. Then, the coalescence was initiated by gradually introducing fluid into the oil droplet, whose size gradually increases until the coalescence with the water droplet, as shown in figure \ref{fig:01}a. The introduction of the oil droplet was by pumping oil through a hole in the substrate (less than 0.1 mm in diameter), following the similar methods of Karpitschka and Riegler (\citeyear{Karpitschka2010SessileCoalescence}). The pre-deposited water droplet was placed at a certain distance from the hole to ensure approximately equal base radii of the two droplets. The typical size of the two droplets was kept around 1 mm or smaller to reduce the effect of gravity.

Droplet pairs of different liquids were used to vary the miscibility and wettability, as shown in table \ref{tab:01}. The droplet with higher surface tension is the water phase and that with lower surface tension is the oil phase. For the water droplets, we used deionised water and its parameter is kept constant in our experiments. For the oil droplets, we used four kinds of liquids, namely silicone oil, alkane, high alcohols, and brominated oils, which are all immiscible with water, but with different surface tensions and interfacial tensions. In addition, four sets of miscible droplet pairs are also prepared for comparison, in which the oil droplets are produced by mixing different ratios of water with ethanol (shown in table \ref{tab:01}). In some experiments, Rhodamine B was added to the water droplet to enhance the contrast between the two droplets, and the concentration was lower than 0.1 wt\%, which was tested to have negligible effects on the droplets' physical properties \citep{Lu2020ImpingementDroplet}. Hydrophobic substrates were used in the experiment. The hydrophobic surface was made by soaking soda-lime glass in a hydrophobic solution (MesoPhobic-2000; MesoBioSystem) and then dried in a drying oven. Two kinds of substrates with different hydrophobicity were obtained by applying different soaking times, and the equilibrium water contact angle is $90 \pm 5^\circ$ for 3-minute soaking and $100 \pm 5^\circ$ for 30-minute soaking (as shown in figure \ref{fig:01}b). For the first kind of hydrophobic substrates, the contact angles for the water/oil interface with the substrate measured from the water side are $137.4 \pm 2.1^\circ$ for silicone oils, $129.4 \pm 3.8^\circ$ for the alkane, and $150.2 \pm 3.2^\circ$ for the brominated oils. For the second kind of hydrophobic substrates, the contact angles for the water/oil interface with the substrate measured from the water side are $140.5 \pm 6.1^\circ$ for silicone oils, $136.0 \pm 4.5^\circ$ for the alkane, and $157.8 \pm 2.1^\circ$ for the brominated oils. The coalescence process was recorded using a high-speed camera (Phantom V1612) with a frame rate ranging from 10,000 to 250,000 frames per second (fps). The typical frame exposure time is 3.38 $\upmu$s, and the resolution is 256$\times$128 for the highest frame rate. In addition, a microscopic lens (Navitar Zoom 6000) was used to achieve the close-up views yielding a resolution of up to 6.37 $\upmu$m$/$ pixel. Backlight was applied by using a high-intensity fiber optic lamp along with a diffuser to illuminate the field of view to achieve uniform lighting. Moreover, in some experiments where the water droplet is dyed, the front lighting positioned to the side of the camera was supplemented to illuminate the droplet for clearer colored visualization. To make sure we could capture the whole process, we started recording long before the contact of the two droplets. We then measure the bridge evolution by starting from the moment when we could first detect a discernible increase in the bridge height, which is a common practice in many studies of droplet coalescence processes \citep{Eddi2013DropletGeometry, Sprittles2014inertialcoalescence}. To further minimize the uncertainty, in our experiments, the initial height of the liquid bridge is limited to less than $5\%$ of the oil droplet height. For some cases with an unusually large initial bridge height (e.g., larger than $5\%$ of the oil droplet height, which may be due to the small spatial inaccuracies of the deposition of the second droplet), they were discarded. All the experiments were performed at room temperature and atmospheric pressure, where the room humidity was kept at over 30\% to avoid the electrostatic effect \citep{Yokota2011ConfinedDroplet}.

\begin{table}
\caption{Physical properties of the liquids used in this study at 20 $^\circ$C. The properties of the silicone oils are from Ref. \citep{Ji2021CompoudJetting}, the alkane from Ref. \citep{Goossens2011DropletSpreading}, the high alcohols from Ref.\ \citep{Yu2019SplittingDroplets}, and the brominated oils from Ref. \citep{Rostami2022CapillaryMerging}. EWM is the ethanol-water-mixture, figures in the parentheses represent the mole fraction of ethanol, and their properties are from Ref. \citep{Khattab2012EthanolMixture}}.
\label{tab:01}
\resizebox{\columnwidth}{!}{%
\begin{tabular}{lcccc}
\hline
Liquids                & Density             & Dynamic viscosity     & Surface tension   & Interfacial tension            \\
                       & $\rho$   (kg/m$^3$) & $\mu$   (mPa$\cdot$s) & $\sigma$   (mN/m) & with water $\sigma$ (mN/m)     \\ \hline
Water                  & 1000                & 1.0                   & 72                & -                              \\
1-Octanol              & 812.4               & 7.4                   & 27.5              & 8.4                            \\
1-Decanol              & 831.2               & 12.2                  & 28.5              & 8.6                            \\
1-Undecanol            & 830                 & 17.5                  & 26.6              & 8.8                            \\
n-Dodecane             & 748.7               & 1.0                   & 22.0             & 48.0                             \\
n-Hexadecane           & 770                 & 3.9                   & 22.1             & 44.4                             \\
Silicone oil (5 cSt)   & 913                 & 4.6                   & $18.7 \pm 0.3$              & $38.1 \pm 0.4$         \\
Silicone oil (10 cSt)  & 930                 & 9.3                   & $18.7 \pm 0.3$              & $38.1 \pm 0.5$          \\
Silicone oil (20 cSt)  & 950                 & 18.9                  & $19.8 \pm 0.7$              & $40.9 \pm 0.5$          \\
Silicone oil (50 cSt)  & 960                 & 48.0                  & $20.4 \pm 0.2$              & $40.9 \pm 0.5$          \\
Silicone oil (100 cSt) & 960                 & 96.0                  & $20.1 \pm 0.2$              & $43.7 \pm 0.4$          \\
Silicone oil (500 cSt) & 970                 & 465                   & $20.0 \pm 0.1$              & $38.7 \pm 0.2$          \\
Bromocyclopentane      & 1386               & $1.4 \pm 0.05$           & $33.2 \pm 0.1$              & 26                    \\
Bromocyclohexane       & 1336               & $2.2 \pm 0.05$           & $32.1 \pm 0.1$              & 26               \\
EWM (1.000)	           & 791.0	            & 1.189	                & 22.85	        & -                             \\
EWM (0.735)	           & 832.9	            & 1.950                 & 24.49         & -                               \\
EWM (0.552)	           & 858.9	            & 2.372	                & 26.00	        & -                               \\
EWM (0.419)	           & 884.4	            & 2.791	                & 27.45	         & -                               \\
\end{tabular}%
}
\end{table}

\section{Results and discussion}\label{sec:3}
\subsection{Stages of the coalescence process}\label{sec:31}
We tracked the outer surface of the two coalescing immiscible droplets as well as their contact line, and the resulting image sequences are shown in figure \ref{fig:02} (also see Movie 1 in Supplementary Material). During droplet coalescence, the governing forces involved are capillary force, inertia, and viscous forces but could be dominant in different regions. The timescales which characterize different dynamics, therefore, could vary from around 1 to 2 milliseconds to several minutes \citep{Narhe2008ContactLine}. Based on the timescale and controlling forces (which will be analysed in subsequent sections), we divided the coalescence process of two immiscible droplets into three primary stages, namely (I) the growth of the liquid bridge, (II) the oscillation of the coalescing sessile droplet, and (III) the formation of a partially-engulfed compound sessile droplet (abbreviated to PECSD) and the subsequent retraction. Each stage is dominated by different forces and exhibits different phenomena.

(I) \emph{The growth of the liquid bridge}: The instant of the first contact in the image sequence is defined as $t = 0$, as shown in figure \ref{fig:02}a. After the two droplets contact, a liquid bridge quickly builds up connecting the two droplets, as shown in figure \ref{fig:03}a. The large curvature formed on the liquid-vapor surface in the bridge area induces a large capillary force, which drives the quick growth of the liquid bridge. The growth rate decreases gradually as the bridge curvature reduces and as more liquid is set in motion. When the bridge height is comparable with the height of the oil droplet (see figure \ref{fig:02}c), the further growth of the liquid bridge is very slow, which can be seen from the time stamps below the snapshots. For example, it takes only around 1.32 ms for the bridge to grow to half of the oil droplet height, but it takes more than twice as long to reach the height of the initial oil droplet as shown in figures \ref{fig:02}c and \ref{fig:02}d. When the bridge height increases to the oil droplet height, it indicates the finish of the first stage. Another prominent feature is the three-phase contact line on the surface of the droplets and as shown in figures \ref{fig:01}c and \ref{fig:01}d, there is a net force along the water surface considering the relative magnitude of the three surface tensions and dynamic contact angles around the three-phase contact line. Consequently, one would expect a precursor film of oil to be driven onto the surface of the water \citep{cuttle2021engulfment,sanjay2022precursorfilm} due to the surface tension gradient, which is often observed in three-phase flows. However, the precursor film is believed not to show up in this first stage since we can observe a sharp discontinuous interface in the bridge region as can be seen in figures \ref{fig:01}c and \ref{fig:01}d. In addition, according to \citep{Koldeweij2019Maragoni}, the evolution of the leading edge of the oil film is $L(t){\sim}{\Delta {{\sigma }^{{1}/{2}}}{{t}^{{3}/{4}}}}/{{{(\mu \rho )}^{{1}/{4}}}}$, where $\Delta \sigma $ is the surface tension difference, $\mu $ and $\rho $ are the dynamic viscosity and density of the droplet of higher surface tension, respectively. Based on this relation, the time for the leading edge of an oil precursor film to reach the top of the water droplet is calculated to be as long as 3--4 ms, which exceeds the timescale of liquid bridge evolution (less than 2 ms in the present case). This further indicates that the precursor film would not influence the bridge growth.

In addition, during the growth of the liquid bridge, the surface capillary waves, which are common in the coalescence of miscible droplets on solid surfaces \citep{Eddi2013DropletGeometry, Lee2013Printing, Sykes2020SubstrateWettability}, are also observed in the present study for immiscible droplets (highlighted by the yellow arrows in figure \ref{fig:03}c).

(II) \emph{The oscillation of the coalescing sessile droplet}: Following the first stage, the coalescing droplet oscillates due to the surface tension and the inertia of the droplet. When the two droplets contact, their pressure difference can be estimated to be positive, $\Delta P={{{\sigma }_{{wa}}}}/{{{R}_{{w}}}}-{{{\sigma }_{{oa}}}}/{{{R}_{{o}}}}>0$, where ${{R}_{{w}}}$ and ${{R}_{{o}}}$ are the initial radii of the water and oil droplets, respectively. Therefore the water droplet first moves towards the oil droplet due to the relatively larger capillary pressure in the water droplet. After the accumulation of the inertia, the droplet overshoots even after the equilibrium point. Surface tension, then, causes the droplet to move backward. This process repeats as the interchange between the surface energy and the kinetic energy with some part of the energy being dissipated both in the bulk and in the contact line. The oscillation process is in a much longer timescale compared with the initial stage of liquid bridge growth.

(III) \emph{The formation of PECSD and its subsequent retraction}: The third stage ensues as the droplet oscillation gradually relaxes. Driven by capillary forces and resisted by both inertia and viscous forces, the oil is seen surrounding the water droplet and filling under the water droplet. As a consequence, the water droplet can be displaced from the solid by the oil (see figure \ref{fig:02}e and figure \ref{fig:01}e), forming a PECSD on the solid surface. A clearer illustration of the displacement process can be seen in figure \ref{fig:01}e, which was obtained through different lighting methods. The PECSD then moves and contracts to decrease the contact area due to the minimization of the surface energy in a much longer timescale (see figure \ref{fig:02}f). This slow stage is characterised by the four-phase capillary-driven flow and the retraction of the PECSD.

Since the three stages are controlled by different mechanisms, the dynamics of each stage will be analyzed in the subsequent sections.

\begin{figure}
  \centering
  \includegraphics[scale=0.9]{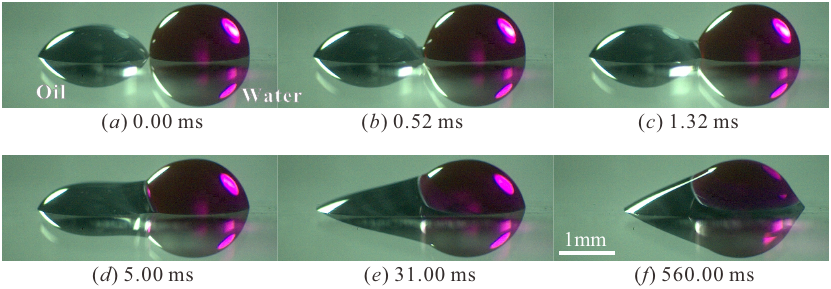}
  \caption{Typical time sequences of coalescence process of immiscible droplets. The two immiscible liquids are water and n-dodecane. The camera was tilted slightly downward (at an angle of 5$^\circ$ to the horizontal) to have a better observation. The volume ratio ${{V}^{*}}={{{V}_{{o}}}}/{{{V}_{{w}}}}$ is 0.5117, where ${V}_{o}$ and ${V}_{w}$ are the volumes of the oil droplet and the water droplet respectively. (Video clip for this process is available in the Supplementary Material: Movie 1).}\label{fig:02}
\end{figure}

\subsection{Growth of liquid bridge}\label{sec:32}
\begin{figure}
  \centering
  \includegraphics[width=\columnwidth]{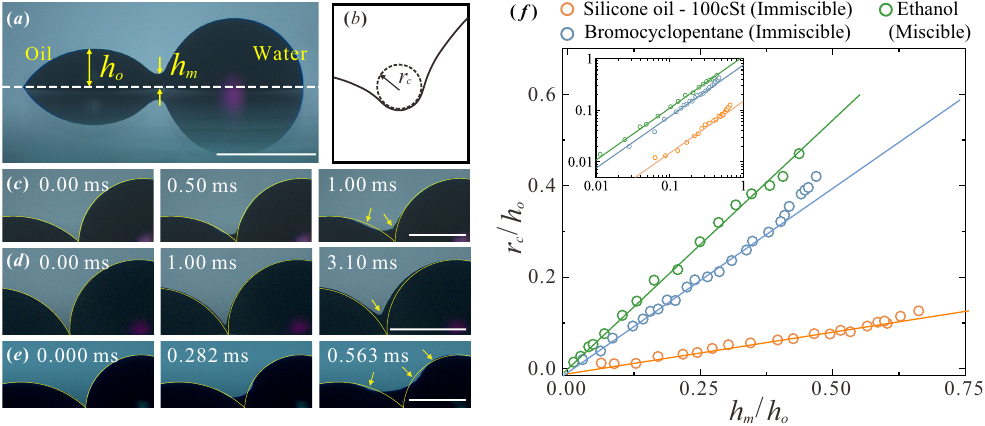}
  \caption{(a) The liquid bridge height ${h}_{m}(t)$ is defined as the vertical distance of the bridge neck to the substrate (white dashed line). b) Illustration of the curvature radius (${r}_{c}$) in the bridge region and is measured by fitting an inscribed circle to the surface shape in the bridge region following the method of \cite{Thoroddsen2007MiscibleDrople}. (c) Snapshots showing the close-up of the liquid bridge of water droplet coalescence with a low-viscosity droplet (bromocyclopentane). The droplet volume ratio ${V}^{*}$ = 0.4315. (d) Snapshots showing the close-up of the liquid bridge of water droplet coalescence with a high-viscosity droplet (silicone oil 100 cSt). The droplet volume ratio ${V}^{*}$ = 0.635. (e) Snapshots showing the close-up of the liquid bridge of water droplet coalescence with an ethanol droplet. The droplet volume ratio ${V}^{*}$ = 0.5894. The overlayed yellow lines on the images are the initial shape of the two droplets. (f) Evolution of curvature radius (${r}_{c}$) with bridge height. The inset shows the data in a log-log plot, where the solid lines have a slope of 1. Both the bridge height and the curvature radius are rescaled with the initial height of the oil droplet. The scale bars (white solid lines) are 1 mm. (Video clips for these processes are available as supplementary material: Movies 2--4).}
\label{fig:03}
\end{figure}

The dynamics of the first stage in the coalescence process is characterised by the rapid growth of the liquid bridge, as shown in figure \ref{fig:03}a. For two sessile droplets on a partial wetting surface, the coalescence of a water droplet with a high-viscosity oil droplet can be different from that of a low-viscosity oil droplet, because the liquid bridge growth could be affected by the extra viscous stress imposed by the substrate and the oil-water-air contact line that is unique to the immiscible liquids. Therefore, the bridge growth is expected to exhibit different behaviours for oil droplets with low viscosity or high viscosity. Here, the high-viscosity droplet and low-viscosity droplet was differentiated by the Ohnesorge number $\Oh={{{\mu }_{o}}}/{\sqrt{{{\rho }_{o}}{{h}_{o}}{{\sigma }_{oa}}}}$, where ${{h}_{o}}$ is the initial height of the oil droplet. Therefore, the low-viscosity oil droplet is identified to have \emph{Oh} in the order of $O(1)$ while the high-viscosity droplet was identified to be $Oh$ much larger than 1.

We first focus on the coalescence of water droplets with low-viscosity oil droplets. For the coalescence of inviscid spherical droplets of identical liquids, the liquid bridge motion has been proved to be driven by capillary forces and resisted by inertia, and this short-time dynamics can be described as the inertia-controlled regime, where the bridge grows as $t^{1/2}$ \citep{Duchemin2003InviscidCoalescence}. While the situation is markedly different in the case of sessile droplets, the evolution of the bridge height ($h_m$ illustrated in figure \ref{fig:03}a) is highly dependent on the droplet geometry before contact. The capillary pressure driving the bridge motion is scaled as ${{P}_{{c}}}\propto {\sigma }/{w}$, and the dynamic pressure in the fluids is estimated as ${{P}_{{i}}}\propto \rho {{\left( {{{h}_{{m}}}}/{t} \right)}^{2}}$, where, for contact angles $\theta < 90^\circ$, the meniscus scale $w$ was found to be proportional to the bridge height \citep{Eddi2013DropletGeometry}: $w\propto {{h}_{{m}}}$. Then, based on the force balance analysis involving the capillary pressure and the dynamic pressure, the bridge growth is predicted to follow ${{h}_{{m}}}\propto {{t}^{{2}/{3}}}$ \citep{Eddi2013DropletGeometry}.

\begin{figure}
  \centering
  \includegraphics[width=\columnwidth]{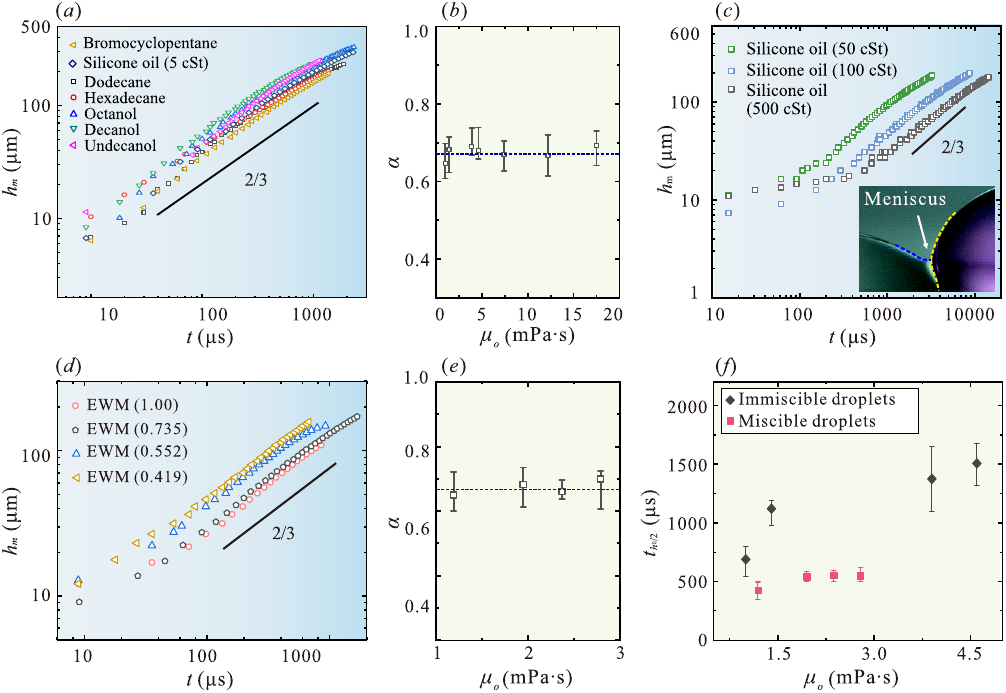}
  \caption{(a) Evolution of the liquid bridge height for water droplets with different low-viscosity oil droplets. The inset shows the data in a linear plot. (b) Fitted exponent $\alpha $ under various oil droplets with different viscosities; the dashed line shows $\alpha  = 2/3$. (c) Evolution of the liquid bridge height for water droplets with different high-viscosity oil droplets. The inset shows the oil meniscus between the oil droplet and the water droplet.  (d) Evolution of the liquid bridge height for water droplets with different miscible droplets made of EWM. (e) Fitted exponent $\alpha $ of the coalescence of miscible droplets with different viscosities. The dashed line shows $\alpha $ = 2/3. (f) Time that it takes for the liquid bridge to reach half of the initial height of the oil droplet.}\label{fig:04}
\end{figure}

In this study of immiscible droplet coalescence, we can see from figure \ref{fig:03}b that the bridge region mainly appears on the lower-surface-tension droplet (i.e., oil), and the bridge height exhibits an almost consistent power-law growth ${{h}_{{m}}}\propto {{t}^{\alpha }}$ as shown in figure \ref{fig:04}a. By referring to the fitting method adopted by \cite{Dekker2022ElasticityCoalescence}, the exponent was obtained by conducting linear fitting in the log-log plot (to determine the first contact instant accurately, the time was slightly shifted until the fitting can achieve the smallest residual sum of squares).
We can see that the exponents are all distributed around 2/3 (see figure \ref{fig:04}b). The rapid bridge growth is driven by the large capillary pressure in the bridge region induced by the large curvature. We then trace the radius of curvature during the bridge growth as shown in figure \ref{fig:03}f, and we can see that the curvature follows approximately a linear growth with the bridge height for both the miscible droplet pairs and immiscible droplet pairs. This linear relationship indicates that the immiscible interface does not change the evolving geometrical relationship between the bridge curvature and the bridge height. When incorporating (${{r}_{c}}\propto {{h}_{m}}$) into the balance between capillary pressure ${{P}_{{c}}}\propto {\sigma }/{w}$ and dynamic pressure ${{P}_{{i}}}\propto \rho {{\left( {{{h}_{{m}}}}/{t} \right)}^{2}}$, it gives the same 2/3 power law of bridge growth of immiscible droplets. However, when comparing the time that it takes for the liquid bridge height to reach half of the initial height of the oil droplet (see figure \ref{fig:04}f), the time for immiscible droplets could be two to three times longer than that of miscible droplets of similar low viscosity. The comparatively longer bridge growth time of immiscible droplets indicates that the growth speed is lower. This lower growth speed can be attributed to the lower kinetic energy converted from the released surface tension energy during the coalescence, as an extra immiscible interface is formed in the immiscible droplets. This immiscible interface would consume part of the released energy that should be converted to kinetic energy. In contrast, this water-oil interface is absent in the miscible droplet coalescence.

These results show that for two droplets of approximately inviscid fluids, the immiscibility of the two droplets has no significant effect on the growth scaling of the liquid bridge height, which follows the same power law as that of the inertial coalescence of the same droplets, i.e., ${{h}_{{m}}}\propto {{t}^{{2}/{3}}}$, but it can substantially decrease the growth speed of liquid bridges.
This is mainly because the curvature of the bridge is mainly on the lower-surface tension droplet. Therefore, the large capillary force driving the bridge growth is provided by the oil droplet. Though a water-oil interface is formed after the contact of the two droplets, its size is small and it has a certain distance to the position of the minimum bridge height. Hence, the effect of the water-oil interface on the bridge growth is weak. Therefore, the growth of the liquid bridge height follows the ${{h}_{{m}}}\propto {{t}^{{2}/{3}}}$ power law.

\begin{figure}
  \centering
  \includegraphics[scale=0.8]{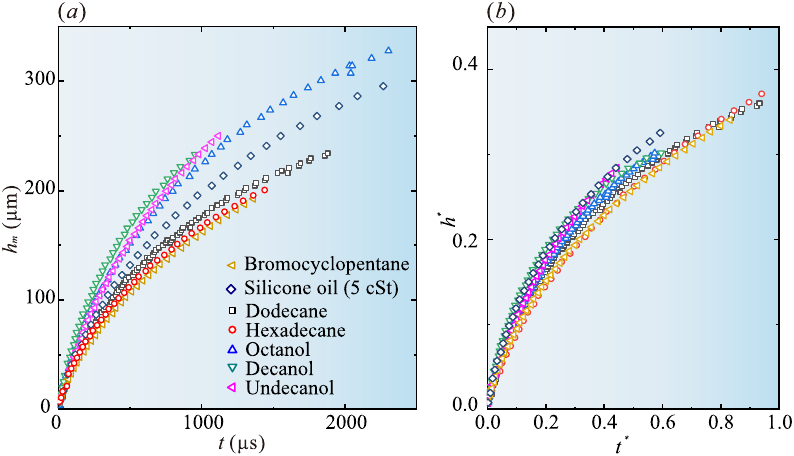}
  \caption{(a) Linear plot of the data in figure \ref{fig:04}a. (b) Rescaling of the data, where the bridge height was scaled by the initial oil droplet height ${h}^{*}$ = ${h}_{m}/{h}_{o}$ , and the time was rescaled with the constructed time scale ${t}^{*}$ = ${t}/{t}_{\sigma}$.}\label{fig:05}
\end{figure}

Further, it is noted that, in the present configuration, the two droplets have different contact angles and, therefore, different sizes. To look into the possible size effect, we first, by assuming the length scale that influences the bridge growth to be the initial oil droplet height ${{h}_{o}}$ , construct a characteristic timescale ${{t}_{\sigma}}=\sqrt{{{{\rho }_\text{avg}}{h}_{o}^{3}}/{\Delta {\sigma } }}$, where ${{\rho }_\text{avg}}={\left( {{\rho }_{o}}+{{\rho }_{w}} \right)}/{2}$ is the average density, and $\Delta {\sigma } ={\sigma }_{{wa}}+{\sigma }_{{oa}}-{\sigma }_{{wo}}$ is based on the net interfacial tension as shown in figure \ref{fig:01}d. The bridge growth data in figure \ref{fig:04}a can then be collapsed together (see figure \ref{fig:05}b), which indicates that the bridge growth is predominantly influenced by the size of the oil droplet compared with the water droplet. It also indicates that the properties of the oil droplets affect the timescale of the bridge growth but do not alter the 2/3 scaling. Similarly, in the experiments of coalescence of asymmetric miscible droplets \citep{Pawar2019AsymmetricCoalescence}, where the contact angle of one droplet is smaller than 90$^\circ$ and is 90$^\circ$ for the other. The bridge growth is found to be dominated by the droplet of contact angle smaller than 90$^\circ$ and to follow the same 2/3 power law but with different prefactors. Here, we do not intend to go further into discussing the exact prefactors but aim to test the power-law dynamics of the bridge growth. In our experiments, the contact radii of the two droplets are limited to be approximately the same to minimize the possible relative size effect.

Moreover, for the coalescence of water droplets with high-viscosity oil droplets (e.g., silicone oils of viscosities ranging from 50 to 500 cSt), the bridge growth is different from that of the low-viscosity droplets in many ways. First, no distinct surface wave is observed on the surface of droplets, while a much sharper curvature in the bridge region is observed, similar to the droplet coalescence scenario of two high-viscosity miscible droplets \citep{Thoroddsen2007MiscibleDrople}.  However, seen from the log-log plot in figure \ref{fig:03}f, the curvature radius also evolve nearly linearly within a period, which may account for the 2/3 power law growth of the bridge. Second, the high viscosity influences the initial growth of the liquid bridge, which leads to a much slower process, as shown in the log-log plot of the data (when $t \ll 1000  {\upmu}$s in figure \ref{fig:04}c). We speculate that this slow motion could be correlated with the extra viscous resistance in the meniscus (see the inset of figure \ref{fig:04}c), which forms when the oil droplet contacts the water droplet. The sharp curvature of the meniscus also indicates large stress in that region, which is reminiscent of the case of a water droplet slowly moving on a viscous-oil-impregnated surface \citep{Keiser2017DropFriction}: during the moving process, the oil meniscus with a sharp curvature was also found surrounding the water droplets and was proposed to be responsible for significant viscous resistance.

\subsection{Oscillation of the coalescing droplet}\label{sec:33}
Due to the liquid velocities induced by the bridge growth and the released surface energy resulting from the coalescence, the droplet surface undergoes further deformation, and then the droplet bulk fluid oscillates. Since the two droplets have different surface tensions and different contact angles, the imbalanced Laplace pressure further generates an obvious horizontal movement. Therefore, we trace the horizontal movement of the droplet centroid, $\Delta X$, in the side-view images by image processing, which could represent the bulk movement of the droplet \citep{Somwanshi2017WallShear, Zhao2021microStriatedSurface}, to investigate the droplet oscillation dynamics.
In the present sessile droplet configuration, the values of Bond number  ${Bo={{\rho} g{{R}^{2}}}/{\sigma}}$, are rather low, being around 0.034 and 0.078 for 1-mm water droplet and oil droplet in air, respectively. According to the analysis of droplet oscillation in Ref. \citep{lamb1924hydrodynamics}, considering the contribution of both surface tension and gravity, the oscillation frequency follows ${f={\sqrt{{g(2l+1)}/{\left[ 2l(l-1)R \right]}+{{\sigma } l(l-1)(l+2)}/{\left( {\rho}{{R}^{3}} \right)}}}/{2\pi }}$, where $l$ is the oscillation degree mode ($l \ge $ 2). The effect of gravity on the oscillation frequency is estimated to be less than 1\%  in our experiments. This indicates that gravitationally forces are less predominant than capillary forces for small droplets, and that capillary forces serve as the dominant restoring force in the present study of oscillation process.

\begin{figure}
  \centering
  \includegraphics[scale=0.9]{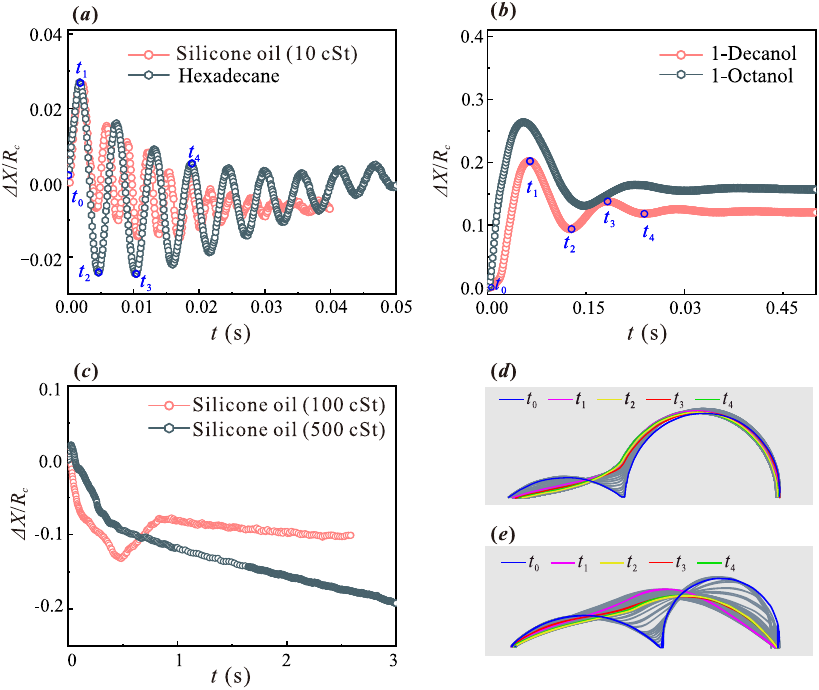}
  \caption{(a-c) Temporal evolution of the normalised horizontal centroid displacement after the coalescence of a water droplet with different oil droplets: (a) silicone oil (10 cSt) and hexadecane; (b) n-decanol and n-octanal; (c) silicone oil (100 cSt) and (500 cSt). $R_c$ is the equivalent radius of the two droplets, ${{R}_{{c}}}=\sqrt[3]{{3\left( {{V}_{{o}}}+{{V}_{{w}}} \right)}/\left({4\pi}\right) }$, $V_o$ and $V_w$ represent the volumes of water droplets and oil droplets, respectively. (d-e) Evolution of the outline of the coalescing droplet: (d) water droplet with hexadecane droplet; (e) water droplet with n-decanol droplet. The corresponding moments are marked in (a) and (b).}\label{fig:06}
\end{figure}

The oscillation for different immiscible droplet pairs shows different behaviours, as shown in figures \ref{fig:06}a-c. For example, in the case of a silicone oil droplet with a water droplet, the oscillation shows a much larger initial displacement amplitude and a lower oscillation period than that when the oil droplet is n-octanol or n-decanol. This different oscillation behaviour should not be attributed to the viscous effect in the low-viscosity liquid, because the viscosity of the 10-cSt silicone oil droplet (9.3 mPa$\cdot$s, in figure \ref{fig:06}a) is even higher than that of the octanol droplet (7.4 mPa$\cdot$s, shown in figure \ref{fig:06}b). Of course, when the oil droplet has a high viscosity, no oscillation behaviour is observed as shown in figure \ref{fig:06}c, which can be explained by the damping effect caused by the high viscosity \citep{Khismatullin2001ShapeOscillation}. However, for the cases shown in figures \ref{fig:06}a and \ref{fig:06}b, the fluid viscosities are always low (3.9--12.2 mPa$\cdot$s). Considering surface tension is the main cause for the oscillation in droplet coalescence dynamics where it is free of external forces \citep{Chashechkin2019dropletShape}, and the liquids used in figures \ref{fig:06}a and \ref{fig:06}b have rather different oil-water interfacial tensions and also different surface tensions, we thus speculate that both the interfacial tension and the surface tension should have a remarkable influence on the droplet oscillation dynamics, and the influence of interfacial tension on the droplet oscillation dynamics should be further investigated. To avoid the damping effect of high viscosity, low-viscosity oil droplets are then used to analyze the oscillation behavior.

\begin{figure}
  \centering
  \includegraphics[scale=0.9]{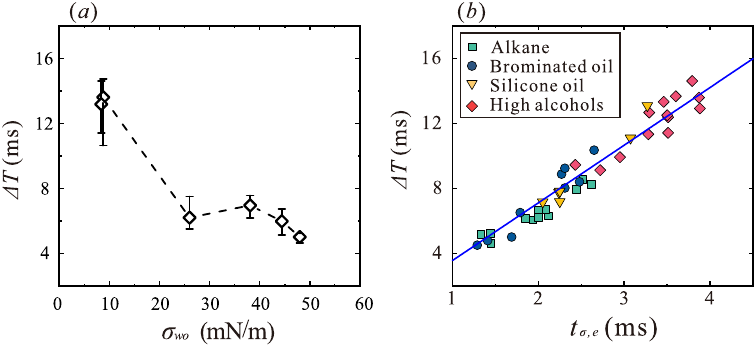}
  \caption{(a) Oscillation period $\Delta T$ of the coalescing immiscible droplets versus the water-oil interfacial tensions. (b) Oscillation period $\Delta T$ versus the capillary timescale ${t}_{\sigma ,e }$ in (\ref{eq:01}).}\label{fig:07}
\end{figure}

We first consider the oscillation process from the perspective of energy conservation. The immiscible water-oil interface would contain a non-negligible portion of the released surface energy which is proportional to the interfacial tension ${{\sigma }_{{wo}}}$, resulting in a decreased energy to be converted to the droplets' kinetic energy given that the viscous dissipation influence is comparably smaller for the low viscosity droplets used. Therefore, the droplet pair with a larger water-oil interfacial tension would end up in a smaller displacement amplitude, as shown in figure \ref{fig:06}a. Further, by plotting the oscillation periods under different interfacial tensions (see figure \ref{fig:07}a), we can see that the oscillation periods show a generally decreasing trend with the interfacial tension. This variation indicates that the interfacial tension, which is unique to the immiscible system, could remarkably affect the coalescence process and lead to different oscillation dynamics. Secondly, the surface tension can quantify an oscillating droplet's ability to restore \citep{Deepu2014Oscillation}, and therefore could also affect the oscillation periods. However, here, both their oil surface tensions and the water-oil interfacial tensions are different, indicating that their frequency differences cannot be attributed to the surface tension or interfacial tension alone. For example, the oscillation period of the water/octanol droplet pair is almost twice as long as that of the water/n-hexadecane droplet pair (these two droplet pairs have different surface tensions and interfacial tensions) as shown in figures \ref{fig:06}a and \ref{fig:06}b. Therefore, both the surface tensions and the interfacial tension should be considered in the analysis of the oscillation dynamics of the coalescing droplets.

Though the role of surface tension in the oscillation of simple droplets has been widely studied \citep{Chireux2015LiquidBridge, Yuan2015LiquidMetal, Khismatullin2001ShapeOscillation, Menchaca2001DropletCoalescence, Zhao2021microStriatedSurface}, the oscillation of the coalescing droplet, which is produced after the coalescence of two immiscible droplets, is more complex. Here, we investigate their oscillation dynamics by considering both their surface tensions and water-oil interfacial tensions. The mass-spring system is commonly used to model the long-time oscillation behaviour of coalescing droplets, during which, surface tension is the restoring force opposing the deformation, and the oscillation of the coalescing droplet can be considered as a capillary-inertia system \citep{Chireux2015LiquidBridge, Yuan2015LiquidMetal}. An immiscible liquid system in capillary-inertia controlled processes (such as droplet impact and oscillation) can be regarded as a parallel spring system \citep{Bernard2020FilmImpact}, and the equivalent spring constant can be represented by the summation of surface tensions and water-oil interfacial tension. With such an analogy, we can first obtain an equivalent spring constant, which is, ${{\sigma }_{{e}}}={{\sigma }_{{wa}}}+{{\sigma }_{{oa}}}+{{\sigma }_{{wo}}}$.

Next, in order to see how the oscillation dynamics is related to the interfacial tension, we evaluate the capillary timescale by taking the equivalent spring constant (i.e., the effective capillary force) into consideration:
\begin{equation}\label{eq:01}
  {{t}_{\sigma ,e}}=\sqrt{\frac{{{m}_{w}}+{{m}_{o}}}{{{\sigma }_{e}}}}=\sqrt{\frac{\pi \left[ \rho_w \left( 3{{r}_{w}}^{2}{{h}_{w}}-{{h}_{w}}^{3} \right)+\rho_o \left( 3{{r}_{o}}^{2}{{h}_{o}}-{{h}_{o}}^{3} \right) \right]}{6{{\sigma }_{{e}}}}},
\end{equation}
where ${{r}_{{w}}}$, ${{h}_{{w}}}$, ${{r}_{{o}}}$, ${{h}_{{o}}}$,  are the initial contact radii and heights of the water and oil droplets, respectively. We then plot the oscillation period versus this proposed capillary timescale ${t}_{\sigma,e}$. As shown in figure \ref{fig:07}b, the oscillation period $\Delta T$ exhibits a good linear relationship with the proposed capillary timescale ${t}_{\sigma,e}$, i.e., $\Delta T\propto {t}_{\sigma,e}$. The linear relationship indicates that ${t}_{\sigma,e }$ properly captures the two immiscible droplets' oscillation dynamics. It also confirms that apart from the surface tensions of the two liquids, the interfacial tension between the two liquids also acts as a non-negligible restoring force which affects the oscillation.

\subsection{Formation and retraction of PECSD}\label{sec:34}
\begin{figure}
  \centering
  \includegraphics[scale=1]{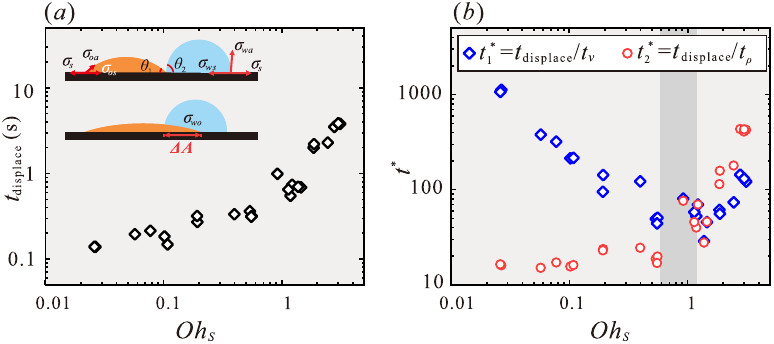}
  \caption{(a) Time taken for the oil droplet to displace the water droplet from the solid substrate ${{t}_{\text{displace}}}$ under different $\Oh_s$. The inset illustrates the interface shapes before and after the displacement process, where ${{\sigma }_{s}}$, ${{\sigma }_{ws}}$, and ${{\sigma }_{os}}$ are the interfacial tensions of solid-air, water-solid, and oil-solid, respectively, and ${{\sigma }_{oa}}$, ${{\sigma }_{wa}}$, and ${{\sigma }_{wo}}$ are the interfacial tensions of oil-air, water-air, and water-oil, respectively. (b) Displacement time ${{t}_{\text{displace}}}$ scaled with an inertia timescale ${{t}_{\rho }}$ (red hollow circles) and a viscous timescale ${{t}_{\nu }}$ (blue hollow prisms).}\label{fig:08}
\end{figure}

After the contact of the two immiscible droplets, in a much longer timescale, the oil is seen spreading below the water droplet driven by the unbalanced capillary force at the liquid-solid interface. Such a capillary-driven flow will eventually lead to the water droplet being displaced by the oil on the substrate, as illustrated by the inset of figure \ref{fig:08}a, which then forms the PECSD. This PECSD then contracts to further lower the system energy.

We analyze the droplet displacing process by referring to the capillary-driven flow in the three-phase system, whose dynamics is characterised by the spreading parameter, $S={{\sigma }_{ij}}-({{\sigma }_{ik}}+{{\sigma }_{jk}})$ \citep{adamson1967physical}. It should be noted that for our four-phase flow (i.e., water, oil, air, and solid), the interfacial tension between the liquid and the solid also influences the spreading dynamics of the droplet \citep{Bonn2009Wetting}, indicating that the wettability of the solid surface should be taken into consideration. We consider this capillary-driven flow from the energetic perspective. When an area $\Delta A$ on the substrate is displaced by the oil (as illustrated in the diagram of figure \ref{fig:08}a), the released surface energy can be written as:
\begin{equation}\label{eq:02}
  \Delta E _\sigma=\Delta A({{\sigma }_{oa}}+{{\sigma }_{ws}})-\Delta A({{\sigma }_{wo}}+{{\sigma }_{os}}),
\end{equation}
where $\Delta A$ is the surface area where the water-solid contact area is replaced by the oil-water and oil-solid surfaces. It should be noted that this is a simplified form, as there is a difference between the area along the substrate $\Delta {{A}}$ and the area along the curved surface of the droplet $\Delta {{A}_{2}}$. The two areas can be correlated with a geometrical relation:
$\Delta {{A}_{2}}={\Delta {A}/{\cos \alpha }}$, where $\alpha $ is the contact angle of oil below the water droplet. Since $\alpha $ is measured to be small (around ${{15}^{\circ}}$), $\cos {\alpha} $ is close to one.
Their respective surface tensions are illustrated in the inset of figure \ref{fig:08}a. The oil-solid and the water-solid interfacial tensions can be obtained from the Young equation, which relates the equilibrium contact angle of the sessile droplet with three surface tensions. For the oil droplet, the oil-solid interfacial tension can be determined as:
\begin{equation}\label{eq:03}
  {{\sigma }_{os}}={{\sigma }_{s}}-{{\sigma }_{oa}}\cos {{\theta }_{1}}.
\end{equation}
For the water droplet, the water-solid interfacial tension can be determined as:
\begin{equation}\label{eq:04}
  {{\sigma }_{ws}}={{\sigma }_{s}}-{{\sigma }_{wa}}\cos {{\theta }_{2}}.
\end{equation}
Substituting (\ref{eq:03}) and (\ref{eq:04}) into (\ref{eq:02}), $\Delta E_\sigma$ is obtained as:
\begin{equation}\label{eq:05}
  \Delta E _\sigma=\Delta A\left[ {{\sigma }_{oa}}(1+\cos {{\theta }_{1}})-{{\sigma }_{wo}}-{{\sigma }_{wa}}\cos {{\theta }_{2}} \right].
\end{equation}
The capillary-driven flow is favored when the released surface energy $\Delta E_\sigma>0$. Inspired by the spreading parameter $S$ which characterises the wetting state of a three-phase system, we here define ${{S}_{\sigma }}={{\sigma }_{oa}}(1+\cos {{\theta }_{1}})-{{\sigma }_{wo}}-{{\sigma }_{wa}}\cos {{\theta }_{2}}$ as the spreading parameter in the present system. In this case, the displacement of water droplets happens under a positive ${{S}_{\sigma }}$, which otherwise would not happen if ${{S}_{\sigma }}$ is negative. We further test this spreading parameter ${{S}_{\sigma }}$ by applying it to the experiments of \cite{Rostami2022CapillaryMerging}, where the merging of two immiscible droplets on solid surfaces was studied. In their experiments, the position of the water droplet was found to show no changes and ${{S}_{\sigma }}$ was calculated to be negative. We, therefore, use ${{S}_{\sigma }}$ to quantify the driving force of this four-phase capillary-driven flow.

We then trace the time taken for the oil droplet to displace the water droplet from the solid surface, ${{t}_{\text{displace}}}$, which is identified from the moment when the two droplets make contact to the moment when the oil reaches the opposite end of the water droplet. During this process, both inertia and viscous forces could resist the flow. The relative importance of these forces can be measured using a modified Ohnesorge number based on the spreading parameter ${{S}_{\sigma }}$:
\begin{equation}\label{eq:06}
  {{\Oh}_{s}}=\frac{{{\mu }_{o}}}{\sqrt{{{\rho }_\text{avg}}{{R}_{c}}{{S}_{\sigma }}}},
\end{equation}
where ${R}_{c}$ is the equivalent radius of the two droplets, ${{R}_{{c}}}=\sqrt[3]{{3\left( {{V}_{{o}}}+{{V}_{{w}}} \right)}/\left({4\pi}\right) }$.

The variation of the displacement time against $\Oh_s$ is shown in figure \ref{fig:08}a. For a small $\Oh_s$, the displacement time does not show significant variation. In contrast, when $\Oh_s >1$, the displacement time is much longer and shows significant variations. Further, based on the spreading parameter ${{S}_{\sigma }}$, two relevant timescales can be involved for the flow from dimensional analysis, a viscous timescale ${{t}_{\nu }}={{{\mu }_{o}}{{R}_{c}}}/{{{S}_{\sigma }}}$ and an inertia timescale ${{t}_{\rho }}=\sqrt{{({{m}_{w}}+{{m}_{o}})}/{{{S}_{\sigma }}}}$. Here, in the inertia timescale, we choose the total mass of the oil droplet and the water droplet $({{m}_{w}}+{{m}_{o}})$ to be the characteristic mass, considering that both the inertia of the two droplets could resist the flow. The measured displacement time is first rescaled with the inertia timescale ${{t}_{\rho }}$, as shown in figure \ref{fig:08}b. The rescaled results show that it is almost a constant for $\Oh_s<1$, but increases with $\Oh_s$ when $\Oh_s >1$. In contrast, by rescaling the data with the viscous timescale ${{t}_{\nu }}$, different behavior can be observed: the dimensionless displacement time shows a significant decrease with $\Oh_s $ when $\Oh_s <1$, and increases when $\Oh_s >1$. The shaded area in figure \ref{fig:08}b, in which $\emph{Oh}_{s}$ is around 1, indicating a transitional regime where the viscous and inertia contributions are comparable. The rescaled result illustrates that for low-viscosity fluids, the droplets' inertia is the main resistance in this capillary-driven flow, while with the increase of $\Oh_s$, both the inertia and viscous forces would have a nonnegligible resistance to this four-phase capillary driven flow.

\begin{figure}
  \centering
  \includegraphics[width=\columnwidth]{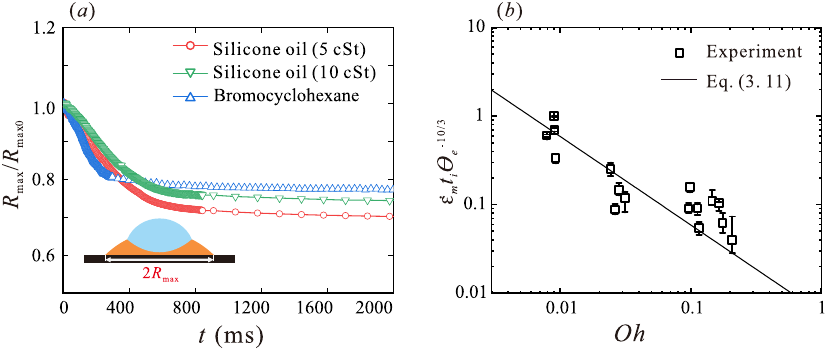}
  \caption{(a) Evolution of the normalised contact radius for three different liquid pairs. The inset illustrates the maximum contact radius $R_{ \max}$ of PECSD during retraction. $R_{{\max}0}$ is the maximum contact radius of PECSD at the outset of retraction. (b) Variation of ${{\dot{\varepsilon }}_{m}}{{t}_{i}}{{\theta }_{e}}^{-{10}/{3}}$ as a function of $\Oh$. The solid line denotes the relationship given in (\ref{eq:11}) with $L$ being the same order as the droplet size $\sim$ 1 mm, and $\lambda $ being the order of nanometer scale $\sim$ 1 nm \citep{Gennes1985WettingStatics, Bonn2009Wetting}.
  }\label{fig:09}
\end{figure}

As the oil displaces the water droplet on the solid surface, a PECSD is formed and is observed to retract at a much slower timescale to further lower the system's energy. Such retraction motion is relatively faster at the beginning of the retraction, and then the retraction speed gradually decreases, as is shown in figure \ref{fig:09}a. Similar retraction processes are common to see in spreading droplets \citep{Andrieu2002SessileCoalescence, Siahcheshm2018DropletRetraction}. A measure of the droplet retraction is the relative retraction rate, defined as $\dot{\varepsilon }={\dot{R}(t)}/{R(t)}$, where $R(t)$ is the contact radius of the compound droplet and the over dot indicates the derivation with respect to time. The retraction rate is a critical parameter to characterise the retraction of sessile droplets \citep{Siahcheshm2018DropletRetraction}.

We then investigate the retraction motion from the energy balance. First, by using the parameters of this PECSD defined before along with the maximum retraction velocity ${u}_{m}$, we can have a quick estimation of the Weber number, $We={{{\rho }_{\text{avg}}}{{R}_{{c}}}{u}_m^{2}}/{\sigma _{{e}}}$, which is in the order of $\emph{O}(10^{-5}$). In this case, due to the low Weber number of this retraction process, the inertia can be safely neglected and the rate of work done by capillary forces can be assumed to be equal to the rate of viscous dissipation \citep{Bonn2009Wetting}. Further, with a low capillary number, the viscous dissipation is concentrated in the region close to the contact line \citep{Gennes1985WettingStatics}. By considering the flow near the contact line of the retracting droplet, the rate of viscous dissipation in the receding corner per unit length of the contact line is expressed as \citep{de2004capillarity}:
\begin{equation}\label{eq:07}
  {{\Phi }_{d}}=\frac{3{{\mu }_{o}}{{U}^{2}}}{{{\theta }_{r}}}\int_{0}^{\infty }{\frac{dx}{x}},
\end{equation}
where $U=\dot{R}(t)$ is the retraction velocity. Since the retraction mainly happens in the direction of the major axis, ${R}(t)$ is defined to be the time-dependent contact radius of the compound droplet measured in the direction of the major axis.
 According to de Gennes' approximation: $\int_{0}^{\infty }{\frac{dx}{x} }\approx\int_{\lambda }^{L}{\frac{dx}{x}}=\ln \left( {L}/{\lambda } \right)$, where the macroscopic and microscopic cutoff lengths $L$ and $\lambda$ are calculated as the droplet size and the molecular size, respectively \citep{Gennes1985WettingStatics}.

The work done by the capillary force per unit length of the contact line can be then expressed as:
\begin{equation}\label{eq:08}
  {{\Phi }_{f}}={{\sigma }_{oa}}\left[ \cos {{\theta }_{r}}(t)-\cos {{\theta }_{r,e}} \right]U,
\end{equation}
where ${{\theta }_{r}}(t)$ is the dynamic receding contact angle, and ${{\theta }_{r,e}}$ is the equilibrium receding contact angle and can be simplified as the equilibrium contact angle. Such simplification is valid in this slow process as the capillary number $Ca={{{\mu }_{{o}}}U}/{{{\sigma }_{{oa}}}}$ is estimated to be in the order of $\emph{O}(10^{-4})$, being much smaller than 1
  \citep{edwards2016dewetting}. Further, in the limit of ${{\theta }_{r}}<{{\theta }_{e}}\ll 1$, a first-order approximation of the Taylor expansion is used $\cos {{\theta }_{r}}(t)\approx 1-{{{\theta }_{r}^{2}}}/{2}$, $\cos {{\theta }_{e}}(t)\approx 1-{{{\theta }_{e}^{2}}}/{2}$, and then (\ref{eq:08}) can be rewritten as ${{\Phi }_{f}}={{{\sigma }_{oa}}[{{\theta }_{e}^{2}}-{{\theta }_{r}^{2}}{{(t)}}]U}/{2}$. The balance between ${{\Phi }_{f}}$ and ${{\Phi }_{d}}$ then gives:
\begin{equation}\label{eq:09}
  \frac{\dot{R}(t)}{R(t)}=\frac{{{\sigma }_{oa}}}{6{{\mu }_{o}}\ln{L}/{\lambda }}{{\left( \frac{\pi }{3{{V}_{t}}} \right)}^{{1}/{3}}}\left[ {{\theta }_{e}^{2}}-{{\theta }_{r}^{2}}{(t)} \right]{{\theta }_{r}^{{4}/{3}}}{{(t)}},
\end{equation}
where ${{V}_{t}}$ is the total volume of the coalescing droplet, which can be estimated by approximating it as a spherical cap, i.e., ${{V}_{t}}={\pi {{\theta }_{r}}(t)R^{3}{{(t)}}}/{4}$. Then, we can obtain the maximum retraction rate by finding ${d\left[ {\dot{R}(t)}/{R(t)} \right]}/{dt}=0$,
\begin{equation}\label{eq:10}
  {{\dot{\varepsilon }}_{m}}=\max \left[ \frac{\dot{R}(t)}{R(t)} \right]=\frac{{{\sigma }_{oa}}}{10{{\mu }_{o}}\ln \left( {L}/{\lambda } \right)}{{\left( \frac{\pi }{3{{V}_{t}}} \right)}^{{1}/{3}}}{{\left( \frac{2}{5} \right)}^{{2}/{3}}}{{\theta }_{e}}^{{10}/{3}}.
\end{equation}
By rearranging (\ref{eq:10}), a relationship for the maximum retraction rate can be obtained:
\begin{equation}\label{eq:11}
  {{\dot{\varepsilon }}_{m}}{{t}_{i}}{{\theta }_{e}}^{{-10}/{3}}=\frac{1}{10\ln \left( {L}/{\lambda } \right)}{{\left( \frac{1}{25} \right)}^{{1}/{3}}}{{\Oh}^{-1}},
\end{equation}
where ${{t}_{i}}=\sqrt{{{{\rho }_\text{avg}}{R}_{c}^{3}}/{{\sigma }_{oa}}}$ is an inertia capillary timescale, and $\Oh={{{\mu }_{o}}}/{\sqrt{{{\rho }_\text{avg}}{{R}_{c}}{{\sigma }_{oa}}}}$ is a modified Ohnesorge number for the retraction process.

Comparing this theoretical model in (\ref{eq:11}) with the experimental data, we can see a good agreement as shown in figure \ref{fig:09}b, confirming the above-proposed mechanism that the retraction of the PECSD is dominated by the viscous dissipation occurring in the receding wedge of the contact line.

\section{Conclusions}\label{sec:4}
In this article, we have investigated the coalescence of two sessile droplets of immiscible fluids on a partial wetting surface. The coalescence process is divided into three stages, namely (I) the growth of the liquid bridge, (II) the oscillation of the coalescing sessile droplet, and (III) the formation of a PECSD and the subsequent retraction. Stage I characterises the rapid growth of the liquid bridge. For low-viscosity oil droplets, the bridge dynamics follows a similar power-law growth as that of miscible droplets. When the oil droplet viscosity increases, we observe a much slower bridge growth in the early time of the first stage, which could be attributed to the extra viscous resistance in the meniscus formed near the oil-water-air contact line. In Stage II, the coalescing sessile droplet oscillates. For high-viscosity oil droplets, the oscillation could be quickly damped out, while for low-viscosity droplets, the oscillation is controlled by inertia and capillary forces. By defining a modified capillary timescale, we show that the interfacial tension between the two immiscible liquids functions as a nonnegligible resistance to the oscillation. In Stage III, a PECSD first forms and then retracts. The formation of the PECSD is characterised by the displacement of water by the oil droplet, and could be resisted by viscous forces or inertia depending on a modified Ohnesorge number $\Oh_s$, which incorporates the influence of the solid surface wettability. To quantify the retraction rate, a model is proposed based on the energy balance, and it demonstrates that the viscous resistance concentrated in a region close to the contact line dominates the slow retraction process.

We expect that the results and insights derived here could be helpful to applications involving microdroplet manipulation, droplet-based biochemical reactions, and liquid removal. To realise better control of droplets in practical applications, external forces can also be adopted, such as magnetic fields \citep{shyam2022magnetofluidic, Li2020ProgrammableDroplets} and electric fields \citep{anand2019electrocoalescence}, which require further study in this area.

\section*{Supplementary data}{\label{SupMat}
Supplementary movies are available online.
\begin{itemize}
  \item Movie 1: Coalescence process of two immiscible droplets, corresponding to figure \ref{fig:02}. The two immiscible liquids are water (right) and n-dodecane (left).
  \item Movie 2: Close-up view of the liquid bridge for the coalescence of a water droplet (right) and a low-viscosity oil droplet (bromocyclopentane, left), corresponding to figure \ref{fig:03}c.
  \item Movie 3: Close-up view of the liquid bridge for the coalescence of a water droplet (right) and a high-viscosity oil droplet (silicone oil 100 cSt, left) , corresponding to figure \ref{fig:03}d.
  \item Movie 4: Close-up view of the liquid bridge for the coalescence of a water droplet (right) and an ethanol droplet (left), corresponding to figure \ref{fig:03}e.
\end{itemize}

\section*{Funding}
This work was supported by the National Natural Science Foundation of China (Grant Nos. 52176083 and 51921004).
\section*{Declaration of interests}
The authors report no conflict of interest.
\section*{Author ORCID}
Zhizhao Che, https://orcid.org/0000-0002-0682-0603

%

\bibliographystyle{jfm}
\bibliography{sessileDroplet}

\end{document}